\documentclass[prd,twocolumn,preprintnumbers,superscriptaddress,nofootinbib,floatfix]{revtex4}
\usepackage[a4paper, hdivide={1.91cm,,1.165cm}, vdivide={1.83cm,,3.6cm}]{geometry}

\usepackage{amstext,amssymb}
\usepackage{amsmath}
\usepackage{graphicx}
\usepackage[hyperfootnotes=false]{hyperref}
\usepackage{xspace}
\usepackage{color}
\usepackage{units}
\usepackage{slashed} 

\begin{document}

\title{The effect of right-handed currents and dark side of the solar neutrino parameter space to Neutrinoless Double Beta Decay}
\author{Pritam Kumar Bishee}
\email{pritamkumarbishee@gmail.com}
\affiliation{Integrated M.Sc. $3^{rd}$ Year, National Institute of Technology, Rourkela, Odisha, India}
\author{Purushottam Sahu}
\email{purushottams@iitbhilai.ac.in}
\author{Sudhanwa Patra}
\email{sudhanwa@iitbhilai.ac.in}
\affiliation{Indian Institute of Technology Bhilai, Raipur 492015, Chhattisgarh, India}
\begin{abstract}
We discuss how dark side of the solar neutrino parameter space and effect of new physics contributions from right-handed currents can reveal the Majorana nature of neutrinos by the observation of the rare process called neutrinoless double beta decay, i.e. the simultaneous decay of two neutrons in the nucleus
of an isotope (A, Z) into two protons and two electrons without the emission of any neutrinos i.e.$(A, Z) \to (A, Z + 2) + 2 e^-$. 
While the standard mechanism of neutrinoless double beta decay with exchange of light Majorana neutrinos, normal ordering and inverted ordering can not saturate the present experimental limit, and quasi-degenerate light neutrinos are strongly disfavored by cosmology, we consider new physics contributions due to  right-handed charged current effects arising in TeV scale left-right symmetric model which can satuare the experimental bound provided by KamLAND-Zen and GERDA.

\end{abstract}

\pacs{} 
\maketitle 
\section{Introduction} 
The recent neutrino oscillation experiments revealed that neutrinos have non-zero masses and mixing, 
which calls for new physics beyond the standard model as the standard model of particle physics predicts massless neutrinos. 
On the other hand, neutrinoless double beta decay($0\nu\beta\beta$) is a unique phenomenon whose experimental observation would reveal whether neutrinos are Majorana particles~\cite{Majorana:1937vz} which violates Lepton Number. 
Majorana particles are its own antiparticles. All the fermions present in the standard model are of Dirac type and only the neutrino, being neutral, can become the Majorana particle. The main parameter of $0\nu\beta\beta$ is effective Majorana mass($m_{ee}$) that depends upon the absolute mass and mass ordering of the neutrino i.e. whether the neutrino mass follows normal ordering (NO) in which third mass eigenstate is the heaviest or inverted ordering (IO) in which the third mass eigenstate is lightest. 
This process has also the potential to tell us about the absolute mass scale and mass ordering of neutrino. So Scientists around the world give their enormous effort and different experiments are going on to tell us about $0\nu\beta\beta$ process. There is no positive signal has been observed yet in any of the experiment.But lower limit of the half-life ($T_{1/2}$) on neutrinoless double beta decay of different isotope i.e. $T_{1/2}(Xe^{156})>1.5\times10^{25}$yrs from KAMLAND-Zen\cite{PhysRevLett.117.082503}, $T_{1/2}(Ge^{76})>8\times10^{25}$yrs from GERDA\cite{PhysRevLett.120.132503} and $T_{1/2}(Te^{130})>1.5\times10^{25}$ yrs from the combined result of CURCINO \& CUORE\cite{PhysRevLett.120.132501} has been found out with $90\%$ C.L .

The discovery of neutrino oscillation, which gives the evidence of neutrino masses and mixing that have an enormous impact on our perception about the understanding of the universe. In the study of particle physics, the most successful and well-accepted theory is the standard model (SM) of the particle physics that has been found to agree with almost all experimental data up to current accelerator energy. All of the particle present in the SM has been experimentally observed. Despite the immense success of the SM, it fails to explain some of the fundamental questions like non zero neutrino mass, the mystery of dark matter and matter-antimatter asymmetry of the universe. So we have to think beyond the standard model (BSM) of particle physics to explain all the shortcoming of SM.\\
A well-motivated candidates of physics beyond the standard model is Left-Right Symmetric Model (LRSM)\cite{PhysRevD.11.2558,PhysRevD.12.1502,PhysRevLett.44.912,PhysRevD.23.165,Senjanovic:1978ev,Heeck:2015qra,Keung:1983uu,Bertolini:2014sua,PhysRevLett.48.848}.It contains additional right-handed current compare to the SM. It explains light neutrino mass via seesaw mechanism by normally adding right-handed neutrino to the model that absent in the SM of particle physics. It also provides the theoretical origin of maximal parity violation that observed in weak interaction while conserved in strong and electromagnetic interaction. It is based on the gauge group $SU(3)_C \times SU(2)_L \times SU(2)_R \times U(1)_{B-L}$. Here the right-handed neutrino is the necessary part of the model. Neutrino acquire their mass from both Type-I and Type-II seesaw mechanism that arises naturally in the LRSM.\\

Neutrino oscillation experiments provide information about mass squared differences and mixing angles i.e. ($\Delta m^2,sin^2 2\theta$). Here we always use ($0 \leq \theta \leq \pi/4$), which is called the "light side" of the parameter space. However, we misses the other half of the parameter space ($\pi/4 < \theta \leq \pi/2$), which is called "dark side"\cite{deGouvea:2000pqg}. Neutrino oscillation in vacuum depends on  $sin^22\theta$, which is equivalent for both light and dark side of the parameter space. That's why we only use the light side of the parameter space. But in the case of matter effect i.e. non-standard neutrino interactions (NSI)\cite{PhysRevD.96.075023,Bakhti2016}, the dark side and the light side are physically inequivalent. The light side solution to the solar neutrino problem generally called standard large mixing angle i.e LMA solution, whereas the dark side solution to the solar neutrino problem called as Dark-LMA i.e DLMA solution.

In this paper, we study the effect of DLMA solution to the solar neutrino problem on neutrinoless double beta ($0\nu\beta\beta$) decay for both of these standard (reprodcued the results presented in ref\,\cite{N.:2019cot}) and extend our new analysis to right-handed current mechanisms and compare them with the standard LMA solution to the solar neutrino problem on $0\nu\beta\beta$ for both mechanisms. This knowledge helps the future experiment to probe in the different energy range of effective mass and find out the sensitivity on $0\nu\beta\beta$.

\section{Standard Mechanism of Neutrinoless Double Beta Decay}
Standard model of particle physics based on the gauge group $SU(3)_C \times SU(2)_L \times U(1)_Y$. It contains only left-handed neutrino. There is absent of right-handed neutrino in the standard model of particle physics. Different experiment are found out the half-life of different isotope for $0\nu\beta\beta$. So for standard mechanism, the inverse half-life ($T_{1/2}$) for $0\nu\beta\beta$ is given as (for earlier references, see \cite{Chakrabortty:2012mh, Awasthi:2013ff,N.:2019cot,Ge:2015yqa,PhysRevLett.106.151801,Pritimita:2016fgr})
\begin{equation}
\centering
[T_{1/2}]^{-1}=G\left| \frac{M_\nu}{m_{e}} \right|^2|m_{ee}^\nu|^2
\end{equation}
where $G$ is the phase factor, ${M_\nu}$ is the nuclear matrix element, ${m_{e}}$ is the mass of electron and $ m_{ee}^\nu$ is the effective majorana mass. We know the value of $G$ and ${M_\nu}$ of different isotope. So the main parameter of interest in $0\nu\beta\beta$ is the effective majorana mass($ m_{ee}^\nu$) which is the combination of neutrino mass eigenvalues and neutrino mixing matrix element. The effective majorana mass is given by
\begin{equation}\label{2}
\centering
m_{ee}^\nu=\left|\sum_{i=1}^{3} U_{ei}^2m_i \right|
\end{equation}
where $U$ is the unitary PMNS mixing matrix and $m_i$ is the mass eigenvalues.\\
PMNS mixing matrix $U$ is given by
\begin{eqnarray}
U 
=
\begin{pmatrix}
1 & 0 & 0 \\
0 & c_{23} & s_{23} \\
0 & -s_{23} & c_{23}
\end{pmatrix}
\begin{pmatrix}
c_{13} & 0 & s_{13}e^{-i\delta} \\
0 & 1 & 0 \\
-s_{13}e^{i\delta}& 0 & c_{13}
\end{pmatrix}\nonumber\\
\begin{pmatrix}
c_{12} & s_{12} & 0 \\
-s_{12} & c_{12} & 0 \\
0 & 0 & 1
\end{pmatrix}
\begin{pmatrix}
1 & 0 & 0 \\
0 & e^\frac{i\alpha}{2} & 0 \\
0 & 0 & e^\frac{i\beta}{2}
\end{pmatrix}
\end{eqnarray}
where $c_{ij}=\cos{\theta_{ij}} , s_{ij}=\sin{\theta_{ij}}$,$\delta$ is the CP violation phases and $\alpha,\beta$ are majorana phases.\\
if we put the value of mixing matrix $U$ in eq.\ref{2} , then effective mass is
\begin{equation}\label{z}
m_{ee}^\nu=\left|m_1c_{12}^2c_{13}^2 + m_2s_{12}^2c_{13}^2e^{i\alpha} + m_3s_{13}^2e^{i\beta} \right|
\end{equation}
Here, the effective mass depends upon the neutrino oscillation parameter $\theta_{12}$,$\theta_{13}$ and the neutrino mass eigenvalues $m_1,m_2$ and $m_3$ for which we don't know the absolute value but we know the mass squared differences between them. and we don't know anything about the majorana phases.\\
we know the value of $\Delta m_{sol}^2$, which is $\Delta m_{sol}^2(\Delta m_{21}^2)=m_2^2-m_1^2$ , has a positive sign so always $m_2>m_1$. and we don't know the sign of $m_{atm}^2(\Delta m_{31}^2)$, which allows two possible ordering of neutrino mass i.e.
\begin{eqnarray}
\Delta m_{atm}^2(\Delta m_{31}^2) &=& m_3^2-m_1^2 ,    \textrm{for Normal Ordering(NO)}\nonumber\\
&=& m_1^2-m_3^2,    \textrm{for Inverted Ordering(IO)}\nonumber
\end{eqnarray}
\underline{\textbf{ Normal Ordering (NO)}} : $m_1<m_2<<m_3$
\begin{eqnarray}
\textrm{Here},\: m_1=m_{\rm{lightest}} \:;\: m_2=\sqrt{m_1^2+\Delta m_{sol}^2}\: ;\nonumber\\  m_3=\sqrt{m_1^2+\Delta m_{sol}^2+\Delta m_{atm}^2}
\end{eqnarray}
\underline{\textbf{ Inverted Ordering (IO)}} : $m_3<<m_1<m_2$
\begin{eqnarray}
\textrm{Here}, \: m_3=m_{\rm{lightest}} \:;\:  m_1=\sqrt{m_3^2+\Delta m_{atm}^2} \:;\nonumber\\  m_2=\sqrt{m_3^2+\Delta m_{atm}^2+\Delta m_{sol}^2}
\end{eqnarray}

\begin{figure}
	\centering
	\includegraphics[width=9cm]{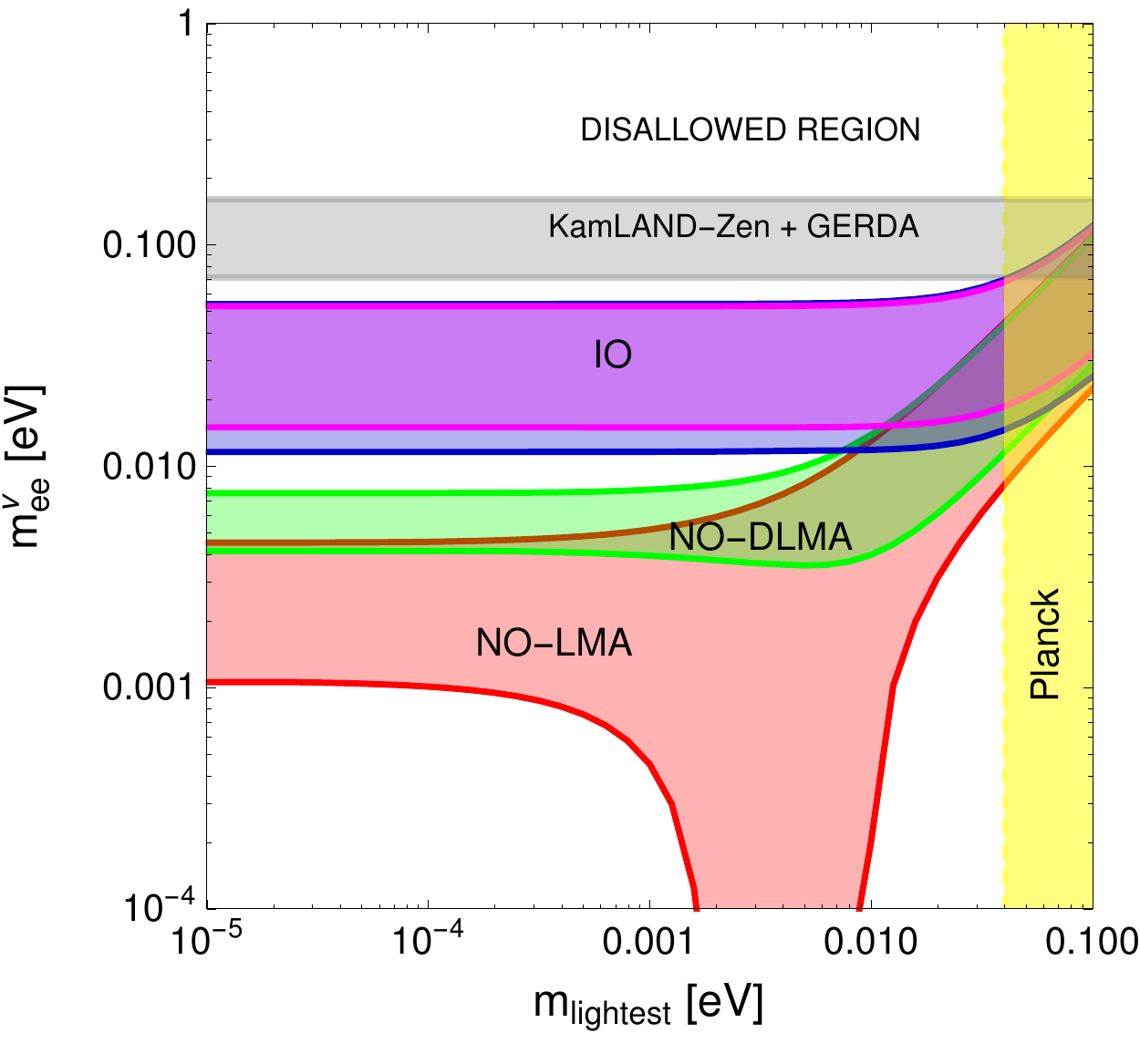}
	\caption{Effective majorana mass $ m_{ee}^\nu$ for neutrinoless double beta decay as a function of lightest neutrino mass for standard mechanism. 
		Here the red and green band are correspond to the solution of $\theta_{12}$ and $\theta_{D12}$ for NO , blue and magenta band are correspond 
		to the solution of $\theta_{12}$ and $\theta_{D12}$ for IO. The analysis has been carried out in standard mechanism of neutrinoless double beta decay in ref\cite{N.:2019cot} and we present here to compare our new results for new physics contributions to neutrinoless double beta decay arising from right-handed current effects.}
	\label{3}
\end{figure}

\begin{table}[htb!]
	\begin{tabular}{cc}
		\hline 
		Oscillation Parameters       & within 3$\sigma$ range   \\
		&  (\cite{deSalas:2017kay})  \\
		\hline \hline
		$\Delta m^2_{\rm {21}} [10^{-5} \mbox{eV}^2]$                & 7.05-8.14   \\
		$|\Delta m^2_{\rm {31}}(\mbox{NO})| [10^{-3} \mbox{eV}^2]$    & 2.41-2.60    \\
		$|\Delta m^2_{\rm {31}}(\mbox{IO})| [10^{-3} \mbox{eV}^2]$   & 2.31-2.51     \\
		\hline
		$\sin^2\theta_{12}$                       & 0.273-0.379 \\
		$\sin^2\theta_{13}(\mbox{NO})$            & 0.0196-0.0241  \\
		$\sin^2\theta_{13}(\mbox{IO})$            & 0.0199-0.0244  \\
		\hline
	\end{tabular}
	\caption{The oscillation parameters like mass squared differences and mixing angles 
		within $3\sigma$ range.\cite{deSalas:2017kay}}
	\label{table-osc}
\end{table}   
In this paper, we symbolize $\theta_{D12}$ for the DLMA solution in presence of NSI and $\theta_{12}$ as the standard LMA solution. The 3$\sigma$ range of both $\theta_{12}$ and $\theta_{D12}$ are given in Table.\ref{1} \cite{deSalas:2017kay, Esteban2018}.
\begin{table}
	\begin{tabular}{ |c|c|c| }
		\hline
		& $sin^2\theta_{12}$ & $sin^2\theta_{D12}$ \\
		\hline
		min  & $0.273$  & $0.650$  \\
		\hline
		max & $0.379$ & $0.725$ \\
		\hline
	\end{tabular}
	\caption{LMA and DLMA solution for $\theta_{12}$ to the solar neutrino problem.}
	\label{1}
\end{table}  
By varying all the neutrino oscillation parameters in their 3$\sigma$ ranges and varied the Majorana phases $\alpha$ and $\beta$ from $0$ to $2\pi$ range, we obtained the plot of effective mass as a function of lightest neutrino mass i.e. $m_1$,$m_3$ for NO and IO respectively in Fig.\ref{3} .

We plot the effective mass by putting LMA and DLMA solution for both NO and IO in Fig.\ref{3}. Here, the gray band($0.07-0.16$ eV) refers to the current upper limit obtained from the combined result of KamLAND-Zen and GERDA\cite{PhysRevLett.120.132503}. The region above this is disallowed. and the yellow region is disallowed by the cosmological constraints on the sum of light neutrino masses\cite{Aghanim:2018eyx}.\\

From the Fig.\ref{3}, For NO, we found out that $ m_{ee}^\nu$ for the DLMA solution is shifting into the region between the NO and IO of LMA solution which is called as the desert region and $ m_{ee}^\nu$ for the DLMA solution is found out to be higher than that of LMA solution. and when the $m_{\rm{lightest}}$ increases, the overlap region between the LMA and DLMA solution also increases. For $m_{\rm{lightest}}\in[10^{-3},10^{-2}]$eV, we found the minimum value of $ m_{ee}^\nu$ for LMA solution is very small i.e. nearly vanishes. but for DLMA solution, in that region minimum value of $ m_{ee}^\nu$ remain same as the value when $m_{\rm{lightest}}<10^{-3}$eV for NO.\\
But in case of IO, the maximum value of $ m_{ee}^\nu$ for both LMA and DLMA solution remain same. whereas the minimum value of $ m_{ee}^\nu$ for DLMA solution is slightly higher than the LMA solution which is nearly same. The overlap region between LMA and DLMA solution remains the same throughout the value of $m_{\rm{lightest}}$. Here The DLMA solution fully overlaps the LMA solution. No considerable change is found out between both of these solutions for IO. So the $ m_{ee}^\nu$ remain same for both LMA and DLMA solution in IO.

\section{Right-handed current effects to Neutrinoless Double Beta Decay}
We believe that lepton number violating $0\nu\beta\beta$ transitions could be induced either by standard mechanism due to the exchange of light Majorana neutrinos discussed in 
previous section or by corresponding new interactions. Since we have already found that standard mechanism can not saturate the present experimental bound one has to 
go beyond SM framework and there exist many models contributing to neutrinoless 
double beta decay \cite{Babu:1995vh,Hirsch:1995vr,Hirsch:1995ek,Hirsch:1996ye,Tello:2010am, Das:2012ii,Barry:2013xxa,Dev:2014iva,Deppisch:2014zta,
	Chakrabortty:2012mh, Awasthi:2013ff, Ge:2015bfa,Patra:2015bga,Deppisch:2014qpa,Patra:2012ur,Nemevsek:2011hz,Awasthi:2015ota,BhupalDev:2013ntw,Deepthi:2019ljo,Ge:2019ldu}.  In the present work, we have considered new interactions 
are arising from purely right-handed currents within left-right symmetric models -- parametrized in terms of the effective mass parameter or 
the half-life of the nucleus -- which can saturate the current experimental bounds and one can derive limits for light Majorana neutrino masses, 
right-handed Majorana neutrinos, right-handed charged gauged boson mass $M_{W_R}$ and its mixing with the left-handed counterpart gauge boson and the corresponding gauge 
coupling $g_R$.

We consider a left-right symmetric model with Type-II seesaw dominance\cite{Ge:2015yqa,PhysRevLett.106.151801,Pritimita:2016fgr} where symmetry breaking occurred at TeV scale leading to right-handed charged gauge boson $W_R$ and right-handed neutrino $N_R$ mass in the order of TeV scale. This leads to new physics contribution to $0\nu\beta\beta$ due to right-handed current via $W_R-W_R$ mediation and heavy neutrino exchange. 

When we considered LRSM with Type-II seesaw dominance, the mass eigenvalues of the left and right handed neutrinos are proportional to each other, 
\begin{equation}
m_L \propto M_R
\end{equation} 
As a result of this, the left and right handed neutrinos have the same PMNS mixing matrix.
\begin{equation}
U_L^{PMNS}=V_R^{PMNS}
\end{equation}
and the mass eigenvalues are related as follows:
\underline{\textbf{ Normal Ordering (NO)}}($m_1=m_{\rm{lightest}}$)
\begin{equation}
m_2=\sqrt{m_1^2+\Delta m_{sol}^2}\: ;\: m_3=\sqrt{m_1^2+\Delta m_{sol}^2+\Delta m_{atm}^2}\: ,\nonumber
\end{equation}
\begin{equation}
M_1=\frac{m_1}{m_3}M_3\:;\:M_2=\frac{m_2}{m_3}M_3
\end{equation}
Here we fixed the heaviest right-handed neutrino mass $M_3$ for NO.\\
\underline{\textbf{ Inverted Ordering (IO)}}($m_3=m_{\rm{lightest}}$)
\begin{equation}
m_1=\sqrt{m_3^2+\Delta m_{atm}^2} \:;\:  m_2=\sqrt{m_3^2+\Delta m_{atm}^2+\Delta m_{sol}^2}\:,\nonumber
\end{equation}
\begin{equation}
M_1=\frac{m_1}{m_2}M_2 \:;\:M_3=\frac{m_3}{m_2}M_2
\end{equation}
Here we fixed the heaviest right-handed neutrino mass $M_2$ for IO.\\
When we considered the LRSM with Type-II see saw dominance where the effect of purely right handed current along with standard mechanism is taken in to consideration, the inverse half-life of a given isotope for $0\nu\beta\beta$ is given by
\begin{eqnarray}
[T_{1/2}]_{LR}^{-1} &=& G\left| \frac{M_\nu}{m_{e}} \right|^2(|m_{ee}^\nu|^2+|m_{ee}^N|^2)\nonumber\\
&=&  G\left| \frac{M_\nu}{m_{e}} \right|^2|m_{ee}^{\nu+N}|^2
\end{eqnarray}
where $m_{ee}^\nu$ is the effective mass arises due to left-handed neutrino in standard mechanism and $m_{ee}^N$ is the effective mass arises due to purely right handed current.
and here $|m_{ee}^{\nu+N}|^2=|m_{ee}^{LR}|^2=|m_{ee}^\nu|^2+|m_{ee}^N|^2$,where $m_{ee}^{LR}$ is the total effective mass arises due to right-handed current in LRSM.\\
We know the expression for $m_{ee}^\nu$ that is given in eq.\ref{z} and the expressions for $m_{ee}^N$ is given by
\begin{eqnarray}
\centering
m_{ee}^N &=& \frac{C_N}{M_3}\left|c_{12}^2c_{13}^2\frac{m_3}{m_1} + s_{12}^2c_{13}^2\frac{m_3}{m_2}e^{i\alpha} + s_{13}^2e^{i\beta} \right|\:\text{(NO)}\nonumber\\
&=& \frac{C_N}{M_2}\left|c_{12}^2c_{13}^2\frac{m_2}{m_1} + s_{12}^2c_{13}^2e^{i\alpha} + s_{13}^2\frac{m_2}{m_3}e^{i\beta} \right|\:\text{(IO)} \nonumber
\end{eqnarray}
\begin{figure}
	\centering
	\includegraphics[width=9cm]{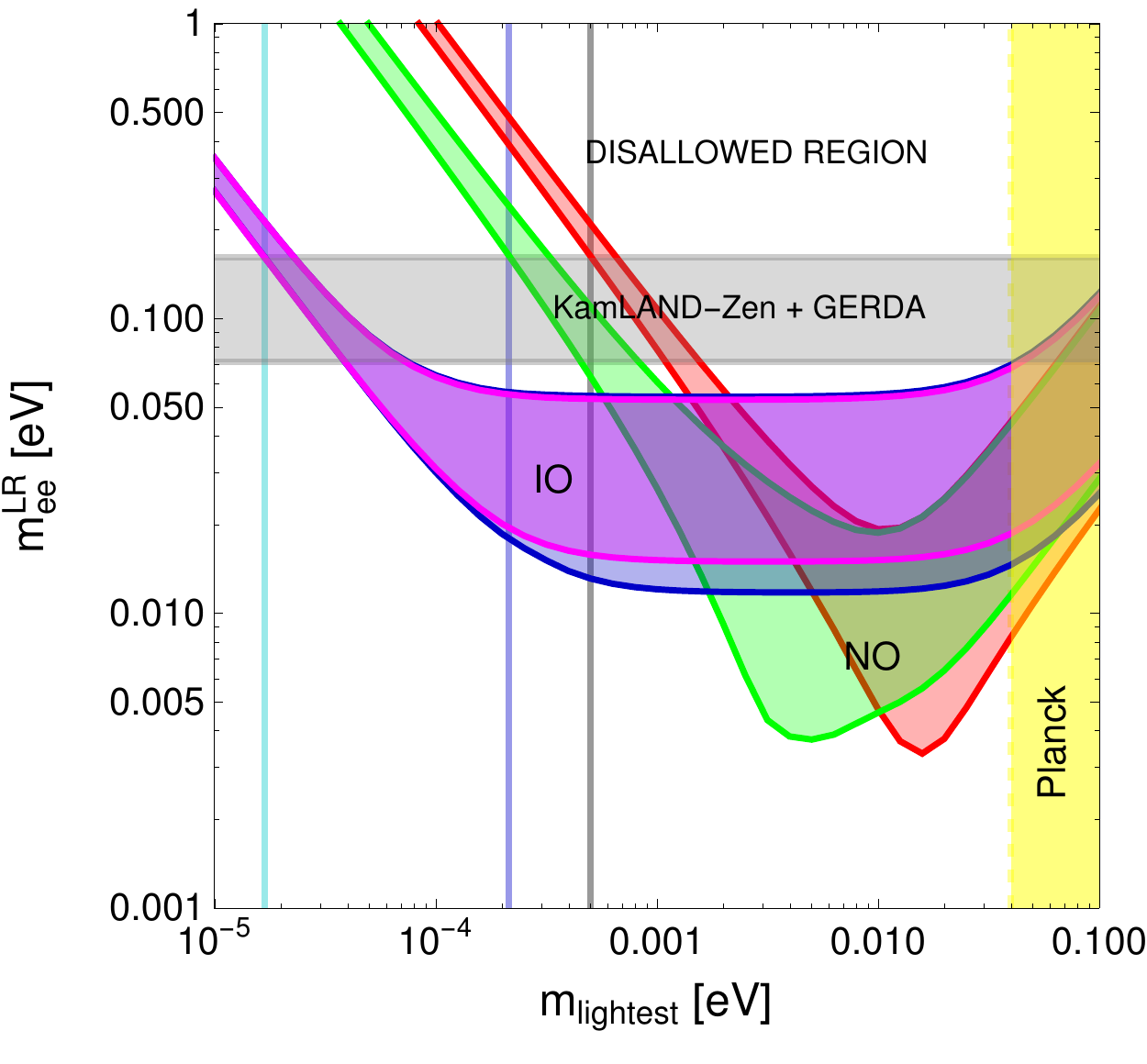}
	\caption{effective majorana mass $ m_{ee}^{LR}$ for neutrinoless double beta decay as a function of lightest neutrino mass from the contribution of right handed current.Here the red and green band are correspond to the solution of $\theta_{12}$ and $\theta_{D12}$ for NO, blue and magenta band are correspond to the solution of $\theta_{12}$ and $\theta_{D12}$ for IO. }
	\label{aa}
\end{figure}
where $C_N=\langle p^2 \rangle  \left(\frac{g_R}{g_L}\right)^4 \left(\frac{M_{W_L}}{M_{W_R}}\right)^4$. Here typical momentum transfer $\langle p \rangle \approx 100$ MeV , $g_R$ \& $g_L$ are the coupling constant of $SU(2)_L$ \& $SU(2)_R$ respectively and $M_{W_L}$ \& $M_{W_R}$ are the mass of the left and right-handed gauge boson i.e $W_L$ and $W_R$ respectively that mediate the process. In this paper, we denote $M_N$ as the heaviest right-handed neutrino mass eigenvalue i.e. $M_3$ for NO and $M_2$ for IO.\\
In the present work, we have considered $g_R \approx g_L$,$M_N=1$ TeV,$M_{W_L}=80.379$ GeV and $M_{W_R}\approx 3.5$ TeV \cite{PhysRevD.98.030001}. By varying all the oscillation parameters in their 3$\sigma$ range\cite{deSalas:2017kay} and Majorana phases from 0 to $2\pi$, we obtained the plot of effective mass ($m_{ee}^{LR}$) as a function of lightest neutrino mass for NO and IO by using both LMA and DLMA solution in Fig.\ref{aa} .\\

From the Fig.\ref{aa}, For NO, we found out that the effective mass($m_{ee}^{LR}$) remain nearly same for the higher value of absolute mass i.e. $m_{\rm{lightest}}>0.03$ eV, but that region is disfavoured by the cosmological constraint. we observed that both of the LMA and DLMA solution for NO saturate the higher value of effective mass limit provided by KamLAND-Zen and GERDA so that we can find the lower limit of absolute mass of the lightest neutrino. Here DLMA solution (Green band) is shifted towards the left as compared with the LMA solution (Red band) that leading to the lower limit of absolute mass for DLMA solution is comparatively smaller than the lower limit of absolute mass for LMA solution. But the $m_{ee}^{LR}$ range is the same for both of these solutions for NO. So the lower limit of absolute mass of the lightest neutrino for NO is found out to be
\begin{equation}
 m_{\rm{lightest}}>5.01\times10^{-4}  \; (\textrm{\;for LMA solution})\nonumber
\end{equation}
\begin{equation}
 m_{\rm{lightest}}>2.15\times10^{-4}  \; (\textrm{for DLMA solution})\nonumber
\end{equation}
In case of IO, $m_{ee}^{LR}$ remain same for both LMA and DLMA solution. So both of these solutions are equal for IO. Here also both of these solutions saturate the higher value of $m_{ee}^{LR}$. As both LMA and DLMA solution are equal for IO, we found the same lower limit of $\rm{lightest}$ for both of these solutions. The lower limit of absolute mass of the lightest neutrino for IO is found to be 
\begin{equation}
m_{\rm{lightest}}>1.685\times10^{-5}\; (\text{IO})\nonumber
\end{equation}
which is very small as compared to the lower limit of absolute mass of lightest neutrino for both of the LMA and DLMA solution for NO.

\section{Conclusion}
If the $0\nu\beta\beta$ process happens, it will address many unsolved fundamental questions of physics like Majorana nature of the neutrino, matter-antimatter asymmetry, absolute mass scale and mass ordering of neutrino that help us for the better understanding of the universe. So searching for this rare process is of paramount importance. From the standard mechanism, we found out that the DLMA solution for NO is shifting into the desert region ($0.004-0.0075$ eV) which provide a new sensitivity goal for the future experiment. If we find out any positive signal for $0\nu\beta\beta$ in that desert region, this will confirm the DLMA solution to the solar neutrino problem as well. From the right-handed current mechanism, we found out that both LMA and DLMA solution to the solar neutrino problem for both NO and IO saturate the experimental limit provided by KamLAND-Zen and GERDA, which also provide the lower limit of absolute mass of lightest neutrino that does not provide by the standard mechanism. So if we find out any positive signal for $0\nu\beta\beta$ above the region of IO in the standard mechanism or in any region that does not explain by the standard mechanism, we have to consider the contribution from the right-handed current as well.

\bibliographystyle{utcaps_mod}
\bibliography{0nubb_DLMANotes}
\end{document}